\documentclass[prl,twocolumn,amssymb,showpacs,amsmath,nobibnotes,
superscriptaddress,floatfix,aps]{revtex4}
\usepackage{graphicx}
\usepackage{bm}

\def\bea{\begin{eqnarray}}
\def\eea{\end{eqnarray}}

\begin{document}
\title{Evidence for Asymptotic Safety from Lattice Quantum Gravity}
\author{J.~Laiho} 
\author{D.~Coumbe}
\affiliation{SUPA, School of Physics and Astronomy, University of Glasgow, Glasgow, G12 8QQ, UK}

\date{\today}
\begin{abstract}
We calculate the spectral dimension for nonperturbative quantum gravity defined via Euclidean dynamical triangulations.  We find that it runs from a value of $\sim3/2$ at short distance to $\sim4$ at large distance scales, similar to results from causal dynamical triangulations.  We argue that the short-distance value of 3/2 for the spectral dimension may resolve the tension between asymptotic safety and the holographic principle.    \end{abstract}

\maketitle




\emph{Introduction}:  The quantization of gravity is one of the great outstanding problems in theoretical physics.  As is well known, a straightforward implementation of general relativity as a perturbative quantum field theory is not renormalizable by power counting, and the divergent counter-terms have been confirmed by explicit calculation \cite{Goroff:1985th}.  Although it is still possible to formulate gravity as an effective field theory at low energies \cite{Donoghue:1997hx}, at each order in the perturbative expansion new divergences appear, requiring an infinite number of couplings that encode the physics at high energy scales, leading to a loss of predictive power.  

Weinberg suggested in \cite{Weinberg:1976xy} that the effective quantum field theory of gravity might be asymptotically safe.  In this scenario, the renormalization group flow of the couplings of the theory would have a nontrivial fixed point, with a finite dimensional ultraviolet critical surface of trajectories attracted to the fixed point at short distances.  Thus, gravity would be ultraviolet complete and describable in terms of a finite, and possibly small, number of parameters, making it effectively renormalizable when formulated nonperturbatively.  

Evidence for the existence of a fixed point in the 4d theory of quantum gravity without matter has come mainly from lattice calculations \cite{Ambjorn:2005qt, Ambjorn:2007jv, Ambjorn:2008wc, Hamber:2009mt} and continuum functional renormalization group methods \cite{Lauscher:2001ya, Lauscher:2001rz, Litim:2003vp, Codello:2007bd, Codello:2008vh, Benedetti:2009rx}.  The renormalization group methods have shown the existence of a nontrivial fixed point in a variety of truncations of the renormalization group equations, indicating that the dimension of the ultraviolet critical surface is finite.  Various truncations with more than three independent couplings suggest that the critical surface is three dimensional \cite{Codello:2008vh, Benedetti:2009rx}. 

 Although the renormalization group studies are suggestive, the truncation of the effective action makes it difficult to systematically assess the reliability of the results obtained using this method.
A lattice formulation of gravity is thus desirable, given the possibility of performing calculations with controlled systematic errors.  In a Euclidean lattice formulation the fixed point would appear as a second order critical point, the approach to which would define a continuum limit \cite{Ambjorn:1997di}.  

There exists an argument due to Banks \cite{Banks:2010tj} (see also Shomer \cite{Shomer:2007vq}) against the possibility of the asymptotic safety scenario.  The argument compares the density of states at high energies expected for a theory of gravity to that of a conformal field theory.  Since a renormalizable quantum field theory is a perturbation of a conformal field theory by relevant operators, a renormalizable field theory must have the same high energy asymptotic density of states as a conformal field theory.  It follows from dimensional analysis, the extensive scaling of the quantities considered, and the fact that a finite-temperature conformal field theory has no dimensionful scales other than the temperature, that the entropy $S$ and energy $E$ scale as
\bea S\sim (RT)^{d-1}, \ \ \ \  E\sim R^{d-1}T^d
\eea
where $R$ is the radius of space-time under consideration, and $T$ is the temperature.  It follows that the entropy of a renormalizable theory must scale with energy as 
\bea\label{eq:CFT}  S\sim E^{\frac{d-1}{d}}.
\eea

For gravity, however, one expects that the high energy spectrum will be dominated by black holes.  The $d$ dimensional 
Schwarzschild solution in asymptotically flat space-time has a black hole with event horizon of radius $r^{d-3}\sim G_N M$, where $G_N$ is Newton's constant and $M$ is the mass of the black hole.  The Bekenstein-Hawking entropy formula tells us that $S\sim r^{d-2}$, so that
\bea\label{eq:GR}  S\sim E^{\frac{d-2}{d-3}}.
\eea
This scaling disagrees with that of Eq.~(\ref{eq:CFT}).  Since considerable evidence supports the holographic argument leading to Eq.~(\ref{eq:GR}), one is led to conclude that gravity cannot be formulated as a renormalizable quantum field theory.  This is a potentially serious obstacle for asymptotic safety, and we revisit this issue below.

Though they do not address this point, lattice calculations with causal dynamical triangulations (CDT) have shown promising results \cite{Ambjorn:2005qt, Ambjorn:2007jv, Ambjorn:2008wc}.  In the parameter space of couplings a phase has been identified with a good semiclassical limit and a ground state consistent with (Euclidean) de Sitter space \cite{Ambjorn:2004pw, Ambjorn:2008wc}.  Also, there exist points connected to this phase that are reasonable candidates for a second order critical point, though establishing the order of the transition at these points remains a difficult numerical problem \cite{Ambjorn:2010hu, Ambjorn:2011ph}.


The original formulation of Euclidean dynamical triangulations (EDT) is a variant of quantum Regge calculus \cite{Ambjorn:1991pq} where the lattice geometries entering the Euclidean path integral are approximated by triangles of fixed edge lengths.  The dynamics is contained in the connectivity of the triangles and their higher dimensional analogs, called simplices.  The dimension of the simplices is fixed, but taking four-dimensional building blocks does not ensure a four-dimensional geometry, and the effective fractal dimension must be determined from the simulations.  There are two phases in the simplest implementation of the model.  Unfortunately, neither of these phases resembles semiclassical general relativity in four dimensions.  Also, the critical point separating these two phases was shown to be first-order, thus ruling out the possibility of taking a continuum limit \cite{Bialas:1996wu, deBakker:1996zx}.

The CDT approach was introduced by Ambjorn and Loll mainly in response to these difficulties \cite{Ambjorn:1998xu}.  The good results obtained so far with the CDT method have been attributed to the introduction of a causality condition, where one distinguishes between spacelike and timelike links on the lattice so that an explicit foliation of the lattice into spacelike hypersurfaces can be introduced, where all the hypersurfaces have the same topology.  Only geometries admitting such a global foliation are included in the space of triangulations defining the measure of the path integral.  In addition to the emergence of de Sitter space \cite{Ambjorn:2008wc}, the CDT calculations lead to the striking result that the effective fractal dimension (the spectral dimension) runs as a function of distance \cite{Ambjorn:2005db}, in accord with renormalization group calculations that support the asymptotic safety scenario \cite{Lauscher:2005qz}.  

Inspired by these successes, we revisit the original EDT formulation in this work.  Our motivations  are the following:
First, the fixing of the foliation in CDT is potentially at odds with general covariance.
If consistent with general covariance, it should be possible to obtain the same results using an EDT formulation, which is explicitly covariant from the outset.  

Second, there are three parameters that enter the bare lattice action in CDT, whereas the original EDT studies typically included only two.  The renormalization group results suggest that the ultraviolet critical surface is three dimensional, so we should consider the possibility that the key ingredient of the CDT formulation is the third coupling, rather than the causality condition.
Both EDT and CDT have the bare Newton and cosmological constants.  CDT also introduces an anisotropy ratio of the lengths of timelike and spacelike links, so that CDT geometries are anisotropic along the time direction \cite{Ambjorn:2008wc}.  Our aim is to study how EDT behaves when a third coupling is introduced. 

\emph{Method}:  Within the EDT formalism, the path integral for Euclidean gravity becomes the discrete partition function \cite{Ambjorn:1991pq, Bilke:1998vj}
\bea\label{eq:Z} Z_{\rm E}  = \sum_{T} \frac{1}{C_T}\left[\prod_{j=1}^{N_2}o(t_j)^{\beta} \right]e^{-S_E},
\eea
where $C_T$ is a symmetry factor that divides out the number of equivalent ways of labeling the vertices in the triangulation $T$.  The Euclidean Einstein-Regge action is
\bea  S_{E}= -\kappa_2 N_2 +\kappa_4 N_4,
\eea
where $N_i$ is the number of simplices of dimension $i$, and where $\kappa_2$ and $\kappa_4$ can be related to the bare Newton's constant $G_N$ and the bare cosmological constant $\Lambda$.  The term in brackets in Eq.~(\ref{eq:Z}) is a nontrivial measure term \cite{Bruegmann:1992jk}, where the product is over all two-simplices (triangles), and $o(t_j)$ is the order of triangle $j$, i.e. the number of four-simplices to which the triangle belongs.  We vary the free parameter $\beta$ as an additional independent coupling constant in the action, bringing the total number of couplings to three.  Most of the previous work on EDT considered the partition function in Eq.~(\ref{eq:Z}) with $\beta=0$ only.


Almost all of the work on four-dimensional EDT used geometries constructed of combinatorial triangulations, in which each d-simplex is defined by a unique set of d+1 distinct vertices, and each d-simplex has a set of unique neighbors.  It is possible to relax this constraint, so that distinct simplices can be defined by the same set of vertices and the neighbors of a d-simplex are no longer unique \cite{Bilke:1998bn}.  It was also shown in Ref.~\cite{Bilke:1998bn} that one must keep the restriction that each four-simplex is defined by a set of unique vertices.  These restrictions define a set of degenerate triangulations.  
It has been shown numerically that using degenerate triangulations leads to a significant reduction in finite-size effects ($\sim$ a factor of 10) for both two-dimensional EDT models, where analytical results are available for comparison \cite{Ambjorn:1997di}, and for four-dimensional models in the unphysical phases where EDT has been studied extensively using combinatorial triangulations \cite{Bilke:1998bn}.  
We use the set of degenerate triangulations in the sum of Eq.~(\ref{eq:Z}), though we also use simulations with combinatorial triangulations as a cross-check.   
    
The numerical implementation of EDT is well-documented, and we refer elsewhere for more details \cite{Ambjorn:1997di}.  In brief, we implement a Monte Carlo integration over a set of four-dimensional degenerate triangulations with the partition function of Eq.~(\ref{eq:Z}) and fixed topology $S^4$.  We use the Metropolis algorithm with a set of local update moves.  It is convenient to keep the four-volume approximately fixed, so we include a term in the action $\delta\lambda|N^f_4-N_4|$ to keep the four-volume close to the value $N^f_4$.  The bare cosmological constant (or equivalently $\kappa_4$) is tuned to its critical value, so that one can take an infinite-volume limit.  This leaves a two dimensional parameter space, which is explored by varying $\kappa_2$ and $\beta$.  Our code is new, but we have checked that we are able to reproduce many results from the extensive literature on combinatorial triangulations and the work of Ref.~\cite{Bilke:1998bn} for degenerate triangulations.

\begin{figure}
\begin{center}
\vspace{-6mm}
\includegraphics[scale=.35]{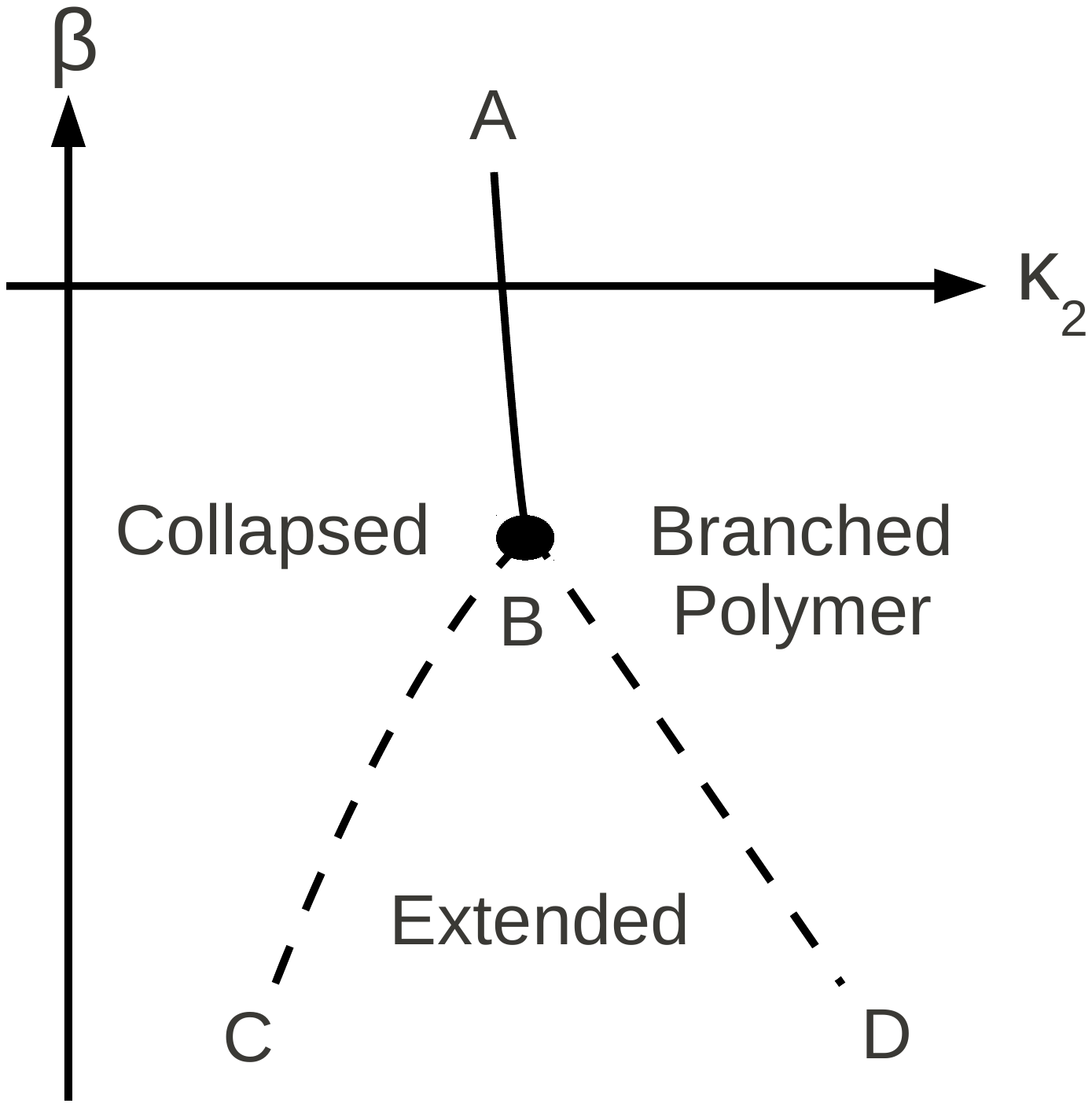}
\vspace{-50mm}
\caption{Schematic of the phase diagram as a function of $\kappa_2$ and $\beta$. \label{fig:phase}}
\end{center}
\vspace{-8mm}
\end{figure}

A schematic illustration of the phase diagram for EDT is shown in Fig.~\ref{fig:phase} within the $\kappa_2$, $\beta$ plane.  The line at $\beta=0$ has received much attention in previous studies, with a first-order phase transition separating the collapsed and branched polymer phases being fairly well-established.  Previous work \cite{Bilke:1998vj} suggests that adding a nontrivial measure term with sufficiently negative $\beta$ leads to the appearance of a third phase, which we label ``extended" in Fig.~\ref{fig:phase}.  Our simulations with degenerate and combinatorial triangulations thus far appear to support this picture.  Although it is nearly certain that line AB is first order, it is not yet clear from our simulations what the order of the transition is along the dotted lines BC and BD in Fig.~\ref{fig:phase}.  The dotted lines indicate that these transitions appear to be softer than the transition across AB, and it is possible that one or both of these may be crossovers.  If this is the case, then the end point of line AB would still be a candidate for a second order critical point.  Determining the order of such phase transitions is difficult, requiring rather large lattices, and the diagram in Fig.~\ref{fig:phase} should be regarded as tentative.  As a first step, we have begun to characterize the extended region to see if it has the behavior expected of a four-dimensional theory of gravity, postponing a detailed study of the critical behavior and the existence of a continuum limit.
  
To determine whether the extended phase has semiclassical gravitational properties, we compute the spectral dimension.  We follow the approach used for CDT in Ref.~\cite{Ambjorn:2005db}.  The geometries that dominate the path-integral are presumably not the smooth manifolds of classical gravity, but rather they are like the paths that dominate the path-integral in nonrelativistic quantum mechanics:  continuous, piecewise linear geometries that are nowhere-differentiable and possess a fractal structure.  The spectral dimension defines the effective dimension of such a fractal geometry via a diffusion process, and it reduces to the usual definition of dimension on smooth manifolds.  


%
%

The spectral dimension of the lattice geometries of the type considered in this work can be obtained from \cite{Ambjorn:2005db}
\bea\label{eq:DS} D_S(\sigma)\approx-2\frac{d  \textrm{log}  \langle P(\sigma) \rangle}{d  \textrm{log}\sigma},
\eea
where $\langle P(\sigma) \rangle$ is the ensemble average of the return probability, i.e. the probability that a random walk will return to its origin after $\sigma$ steps.  This formula receives finite-size corrections when $\sigma$ becomes much larger than the lattice volume $N_4^{2/D_S}$.
The spectral dimension is not the only way to define an effective dimension on a fractal and in general will differ from other definitions of dimension, such as the Hausdorff dimension.  The spectral dimension warrants particular attention because it is the relevant scaling dimension entering the thermodynamic equations of state on a fractal geometry \cite{Akkermans:2010dz}.

 \begin{figure}
  \vspace{-38mm}
\begin{center}
\includegraphics[scale=.44]{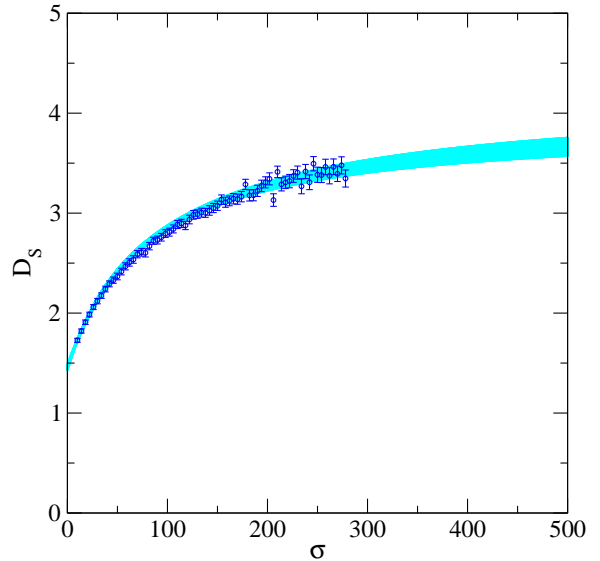}
\vspace{-13mm}
\caption{Spectral dimension as a function of diffusion time $\sigma$, including a fit to Eq.~(\ref{eq:DSfit}).  The width of the band shows the statistical error in the fit.   \label{fig:DS}}
\end{center}
\vspace{-6mm}
\end{figure}

\emph{Results}:  Figure~\ref{fig:DS} shows our new result for the spectral dimension plotted as a function of $\sigma$.  This is the central result of our letter.  We obtained this result from an ensemble of degenerate triangulations with a volume of 4000 four-simplices in the extended phase with $\kappa_2=2.1$ and $\beta=-1.0$.  We analyzed $\sim 1000$ configurations using the average of the return probability as in Eq.~(\ref{eq:DS}), and the errors are determined from a single elimination jackknife procedure.  To reduce statistical errors the data were binned in bin sizes of two from $\sigma=2$ to 80 and in bin sizes of 4 from $\sigma=80$ to 288.  The fit is to the functional form suggested by Ref.~\cite{Ambjorn:2005db},
\bea\label{eq:DSfit}   D_S(\sigma) = a - \frac{b}{c+\sigma},
\eea
with the constants $a$, $b$, and $c$ determined by the fit.  

Our preferred fit is shown in Fig.~\ref{fig:DS} with statistical errors.  This fit uses the full covariance matrix in the estimate of $\chi^2$ with $\sigma$ ranging from 10 to 146 in steps of 4, yielding a $\chi^2/\textrm{dof}=35/32$ and a confidence level (corrected for finite sample size) of CL$=0.37$.  Variations of the fit range and fit function were used to estimate a systematic error on the asymptotic value of $D_S$, always under the assumption that $D_S$ approaches a constant as $\sigma$ goes to infinity.  The results of our preferred fit are $D_S(\infty)=4.04\pm 0.26$,  $D_S(0)=1.457\pm 0.064$, where the errors include both the statistical error and a systematic error associated with varying the fit function and the fit range added in quadrature.  As one can see from the small $\sigma$ data, $D_S(0)$ is not very sensitive to the choice of fit function, nor to changing the minimum value of $\sigma$ to values up to $40$.  The behavior of the spectral dimension is generic within the extended phase and does not depend on the precise values of the bare parameters used in the action.  Finally, we remark that calculations of the spectral dimension with combinatorial triangulations in the extended phase give similar results, but require significantly larger lattice volumes.

\emph{Conclusions}:  Our results for the running spectral dimension are in reasonable agreement with the CDT results: $D_S(\infty)=4.02\pm 0.1$ and $D_S(0)=1.80\pm 0.25$ \cite{Ambjorn:2005db}.  This supports the idea that the explicitly covariant EDT formulation is a valid candidate for a theory of quantum gravity, with a spectral dimension $\sim 4$ at long distances, and it provides nontrivial evidence that CDT and EDT may be in the same universality class.

Our result for the short-distance spectral dimension $D_S(0)=1.457\pm 0.064$ is inconsistent with the renormalization group result that $D_S(0)=2$ exactly \cite{Lauscher:2005qz}.  However, returning to the holographic argument against asymptotic safety, we note that for the asymptotic density of states one should use the short-distance spectral dimension as the scaling dimension in Eq.~(\ref{eq:CFT}).  Under the plausible assumption that the relevant dimension in the holographic scaling argument is also the spectral dimension, we note that Eqs.~(\ref{eq:CFT}) and (\ref{eq:GR}) agree precisely when $D_S=3/2$.  Then, our result from EDT resolves the tension between asymptotic safety and holographic entropy scaling.  In light of this, it is worth revisiting the renormalization group arguments leading to $D_S(0)=2$, as well as embarking on more precise CDT calculations to see whether they confirm the value $3/2$.  If one could demonstrate the existence of a continuum limit at a second order critical point, this result could be the first clue as to how a renormalizable quantum field theory of gravity may yet be consistent with the holographic principle.

We are grateful to C. DeTar, R. Dowdall, D. Miller, and C. White for useful discussions and to A. Kronfeld for comments on the manuscript.  This work was funded by STFC and the Scottish Universities Physics Alliance.  Computing was done on Scotgrid and on the DiRAC facility jointly funded by STFC and BIS.

\bibliography{SuperBib}

\begin{thebibliography}{29}
\expandafter\ifx\csname natexlab\endcsname\relax\def\natexlab#1{#1}\fi
\expandafter\ifx\csname bibnamefont\endcsname\relax
  \def\bibnamefont#1{#1}\fi
\expandafter\ifx\csname bibfnamefont\endcsname\relax
  \def\bibfnamefont#1{#1}\fi
\expandafter\ifx\csname citenamefont\endcsname\relax
  \def\citenamefont#1{#1}\fi
\expandafter\ifx\csname url\endcsname\relax
  \def\url#1{\texttt{#1}}\fi
\expandafter\ifx\csname urlprefix\endcsname\relax\def\urlprefix{URL }\fi
\providecommand{\bibinfo}[2]{#2}
\providecommand{\eprint}[2][]{\url{#2}}

\bibitem[{\citenamefont{Goroff and Sagnotti}(1986)}]{Goroff:1985th}
\bibinfo{author}{\bibfnamefont{M.~H.} \bibnamefont{Goroff}} \bibnamefont{and}
  \bibinfo{author}{\bibfnamefont{A.}~\bibnamefont{Sagnotti}},
  \bibinfo{journal}{Nucl.Phys.} \textbf{\bibinfo{volume}{B266}},
  \bibinfo{pages}{709} (\bibinfo{year}{1986}).

\bibitem[{\citenamefont{Donoghue}(1997)}]{Donoghue:1997hx}
\bibinfo{author}{\bibfnamefont{J.}~\bibnamefont{Donoghue}}
  (\bibinfo{year}{1997}), \eprint{gr-qc/9712070}.

\bibitem[{\citenamefont{Weinberg}(1976)}]{Weinberg:1976xy}
\bibinfo{author}{\bibfnamefont{S.}~\bibnamefont{Weinberg}}
  (\bibinfo{year}{1976}).

\bibitem[{\citenamefont{Ambjorn
  et~al.}(2005{\natexlab{a}})\citenamefont{Ambjorn, Jurkiewicz, and
  Loll}}]{Ambjorn:2005qt}
\bibinfo{author}{\bibfnamefont{J.}~\bibnamefont{Ambjorn}},
  \bibinfo{author}{\bibfnamefont{J.}~\bibnamefont{Jurkiewicz}},
  \bibnamefont{and} \bibinfo{author}{\bibfnamefont{R.}~\bibnamefont{Loll}},
  \bibinfo{journal}{Phys. Rev.} \textbf{\bibinfo{volume}{D72}},
  \bibinfo{pages}{064014} (\bibinfo{year}{2005}{\natexlab{a}}),
  \eprint{hep-th/0505154}.

\bibitem[{\citenamefont{Ambjorn
  et~al.}(2008{\natexlab{a}})\citenamefont{Ambjorn, Gorlich, Jurkiewicz, and
  Loll}}]{Ambjorn:2007jv}
\bibinfo{author}{\bibfnamefont{J.}~\bibnamefont{Ambjorn}},
  \bibinfo{author}{\bibfnamefont{A.}~\bibnamefont{Gorlich}},
  \bibinfo{author}{\bibfnamefont{J.}~\bibnamefont{Jurkiewicz}},
  \bibnamefont{and} \bibinfo{author}{\bibfnamefont{R.}~\bibnamefont{Loll}},
  \bibinfo{journal}{Phys.Rev.Lett.} \textbf{\bibinfo{volume}{100}},
  \bibinfo{pages}{091304} (\bibinfo{year}{2008}{\natexlab{a}}),
  \eprint{arXiv:0712.2485}.

\bibitem[{\citenamefont{Ambjorn
  et~al.}(2008{\natexlab{b}})\citenamefont{Ambjorn, Gorlich, Jurkiewicz, and
  Loll}}]{Ambjorn:2008wc}
\bibinfo{author}{\bibfnamefont{J.}~\bibnamefont{Ambjorn}},
  \bibinfo{author}{\bibfnamefont{A.}~\bibnamefont{Gorlich}},
  \bibinfo{author}{\bibfnamefont{J.}~\bibnamefont{Jurkiewicz}},
  \bibnamefont{and} \bibinfo{author}{\bibfnamefont{R.}~\bibnamefont{Loll}},
  \bibinfo{journal}{Phys. Rev.} \textbf{\bibinfo{volume}{D78}},
  \bibinfo{pages}{063544} (\bibinfo{year}{2008}{\natexlab{b}}),
  \eprint{arXiv:0807.4481}.

\bibitem[{\citenamefont{Hamber}(2009)}]{Hamber:2009mt}
\bibinfo{author}{\bibfnamefont{H.~W.} \bibnamefont{Hamber}},
  \bibinfo{journal}{Gen.Rel.Grav.} \textbf{\bibinfo{volume}{41}},
  \bibinfo{pages}{817} (\bibinfo{year}{2009}), \eprint{arXiv:0901.0964}.

\bibitem[{\citenamefont{Lauscher and
  Reuter}(2002{\natexlab{a}})}]{Lauscher:2001ya}
\bibinfo{author}{\bibfnamefont{O.}~\bibnamefont{Lauscher}} \bibnamefont{and}
  \bibinfo{author}{\bibfnamefont{M.}~\bibnamefont{Reuter}},
  \bibinfo{journal}{Phys.Rev.} \textbf{\bibinfo{volume}{D65}},
  \bibinfo{pages}{025013} (\bibinfo{year}{2002}{\natexlab{a}}),
  \eprint{hep-th/0108040}.

\bibitem[{\citenamefont{Lauscher and
  Reuter}(2002{\natexlab{b}})}]{Lauscher:2001rz}
\bibinfo{author}{\bibfnamefont{O.}~\bibnamefont{Lauscher}} \bibnamefont{and}
  \bibinfo{author}{\bibfnamefont{M.}~\bibnamefont{Reuter}},
  \bibinfo{journal}{Class.Quant.Grav.} \textbf{\bibinfo{volume}{19}},
  \bibinfo{pages}{483} (\bibinfo{year}{2002}{\natexlab{b}}),
  \eprint{hep-th/0110021}.

\bibitem[{\citenamefont{Litim}(2004)}]{Litim:2003vp}
\bibinfo{author}{\bibfnamefont{D.~F.} \bibnamefont{Litim}},
  \bibinfo{journal}{Phys.Rev.Lett.} \textbf{\bibinfo{volume}{92}},
  \bibinfo{pages}{201301} (\bibinfo{year}{2004}), \eprint{hep-th/0312114}.

\bibitem[{\citenamefont{Codello et~al.}(2008)\citenamefont{Codello, Percacci,
  and Rahmede}}]{Codello:2007bd}
\bibinfo{author}{\bibfnamefont{A.}~\bibnamefont{Codello}},
  \bibinfo{author}{\bibfnamefont{R.}~\bibnamefont{Percacci}}, \bibnamefont{and}
  \bibinfo{author}{\bibfnamefont{C.}~\bibnamefont{Rahmede}},
  \bibinfo{journal}{Int.J.Mod.Phys.} \textbf{\bibinfo{volume}{A23}},
  \bibinfo{pages}{143} (\bibinfo{year}{2008}), \eprint{arXiv:0705.1769}.

\bibitem[{\citenamefont{Codello et~al.}(2009)\citenamefont{Codello, Percacci,
  and Rahmede}}]{Codello:2008vh}
\bibinfo{author}{\bibfnamefont{A.}~\bibnamefont{Codello}},
  \bibinfo{author}{\bibfnamefont{R.}~\bibnamefont{Percacci}}, \bibnamefont{and}
  \bibinfo{author}{\bibfnamefont{C.}~\bibnamefont{Rahmede}},
  \bibinfo{journal}{Annals Phys.} \textbf{\bibinfo{volume}{324}},
  \bibinfo{pages}{414} (\bibinfo{year}{2009}), \eprint{arXiv:0805.2909}.

\bibitem[{\citenamefont{Benedetti et~al.}(2009)\citenamefont{Benedetti,
  Machado, and Saueressig}}]{Benedetti:2009rx}
\bibinfo{author}{\bibfnamefont{D.}~\bibnamefont{Benedetti}},
  \bibinfo{author}{\bibfnamefont{P.~F.} \bibnamefont{Machado}},
  \bibnamefont{and}
  \bibinfo{author}{\bibfnamefont{F.}~\bibnamefont{Saueressig}},
  \bibinfo{journal}{Mod.Phys.Lett.} \textbf{\bibinfo{volume}{A24}},
  \bibinfo{pages}{2233} (\bibinfo{year}{2009}), \eprint{arXiv:0901.2984}.

\bibitem[{\citenamefont{Ambjorn et~al.}(1997)\citenamefont{Ambjorn, Durhuus,
  and Jonsson}}]{Ambjorn:1997di}
\bibinfo{author}{\bibfnamefont{J.}~\bibnamefont{Ambjorn}},
  \bibinfo{author}{\bibfnamefont{B.}~\bibnamefont{Durhuus}}, \bibnamefont{and}
  \bibinfo{author}{\bibfnamefont{T.}~\bibnamefont{Jonsson}}
  (\bibinfo{year}{1997}).

\bibitem[{\citenamefont{Banks}(2010)}]{Banks:2010tj}
\bibinfo{author}{\bibfnamefont{T.}~\bibnamefont{Banks}} (\bibinfo{year}{2010}),
  \eprint{arXiv:1007.4001}.

\bibitem[{\citenamefont{Shomer}(2007)}]{Shomer:2007vq}
\bibinfo{author}{\bibfnamefont{A.}~\bibnamefont{Shomer}}
  (\bibinfo{year}{2007}), \eprint{arXiv:0709.3555}.

\bibitem[{\citenamefont{Ambjorn
  et~al.}(2005{\natexlab{b}})\citenamefont{Ambjorn, Jurkiewicz, and
  Loll}}]{Ambjorn:2004pw}
\bibinfo{author}{\bibfnamefont{J.}~\bibnamefont{Ambjorn}},
  \bibinfo{author}{\bibfnamefont{J.}~\bibnamefont{Jurkiewicz}},
  \bibnamefont{and} \bibinfo{author}{\bibfnamefont{R.}~\bibnamefont{Loll}},
  \bibinfo{journal}{Phys.Lett.} \textbf{\bibinfo{volume}{B607}},
  \bibinfo{pages}{205} (\bibinfo{year}{2005}{\natexlab{b}}),
  \eprint{hep-th/0411152}.

\bibitem[{\citenamefont{Ambjorn et~al.}(2010)\citenamefont{Ambjorn, Gorlich,
  Jordan, Jurkiewicz, and Loll}}]{Ambjorn:2010hu}
\bibinfo{author}{\bibfnamefont{J.}~\bibnamefont{Ambjorn}},
  \bibinfo{author}{\bibfnamefont{A.}~\bibnamefont{Gorlich}},
  \bibinfo{author}{\bibfnamefont{S.}~\bibnamefont{Jordan}},
  \bibinfo{author}{\bibfnamefont{J.}~\bibnamefont{Jurkiewicz}},
  \bibnamefont{and} \bibinfo{author}{\bibfnamefont{R.}~\bibnamefont{Loll}},
  \bibinfo{journal}{Phys.Lett.} \textbf{\bibinfo{volume}{B690}},
  \bibinfo{pages}{413} (\bibinfo{year}{2010}), \eprint{arXiv:1002.3298}.

\bibitem[{\citenamefont{Ambjorn et~al.}(2011)\citenamefont{Ambjorn, Gorlich,
  Jurkiewicz, Loll, Gizbert-Studnicki et~al.}}]{Ambjorn:2011ph}
\bibinfo{author}{\bibfnamefont{J.}~\bibnamefont{Ambjorn}},
  \bibinfo{author}{\bibfnamefont{A.}~\bibnamefont{Gorlich}},
  \bibinfo{author}{\bibfnamefont{J.}~\bibnamefont{Jurkiewicz}},
  \bibinfo{author}{\bibfnamefont{R.}~\bibnamefont{Loll}},
  \bibinfo{author}{\bibfnamefont{J.}~\bibnamefont{Gizbert-Studnicki}},
  \bibnamefont{et~al.} (\bibinfo{year}{2011}), \eprint{arXiv:1102.3929}.

\bibitem[{\citenamefont{Ambjorn and Jurkiewicz}(1992)}]{Ambjorn:1991pq}
\bibinfo{author}{\bibfnamefont{J.}~\bibnamefont{Ambjorn}} \bibnamefont{and}
  \bibinfo{author}{\bibfnamefont{J.}~\bibnamefont{Jurkiewicz}},
  \bibinfo{journal}{Phys.Lett.} \textbf{\bibinfo{volume}{B278}},
  \bibinfo{pages}{42} (\bibinfo{year}{1992}), \bibinfo{note}{revision of
  NBI-HE-91-47}.

\bibitem[{\citenamefont{Bialas et~al.}(1996)\citenamefont{Bialas, Burda,
  Krzywicki, and Petersson}}]{Bialas:1996wu}
\bibinfo{author}{\bibfnamefont{P.}~\bibnamefont{Bialas}},
  \bibinfo{author}{\bibfnamefont{Z.}~\bibnamefont{Burda}},
  \bibinfo{author}{\bibfnamefont{A.}~\bibnamefont{Krzywicki}},
  \bibnamefont{and}
  \bibinfo{author}{\bibfnamefont{B.}~\bibnamefont{Petersson}},
  \bibinfo{journal}{Nucl.Phys.} \textbf{\bibinfo{volume}{B472}},
  \bibinfo{pages}{293} (\bibinfo{year}{1996}), \eprint{hep-lat/9601024}.

\bibitem[{\citenamefont{de~Bakker}(1996)}]{deBakker:1996zx}
\bibinfo{author}{\bibfnamefont{B.~V.} \bibnamefont{de~Bakker}},
  \bibinfo{journal}{Phys.Lett.} \textbf{\bibinfo{volume}{B389}},
  \bibinfo{pages}{238} (\bibinfo{year}{1996}), \eprint{hep-lat/9603024}.

\bibitem[{\citenamefont{Ambjorn and Loll}(1998)}]{Ambjorn:1998xu}
\bibinfo{author}{\bibfnamefont{J.}~\bibnamefont{Ambjorn}} \bibnamefont{and}
  \bibinfo{author}{\bibfnamefont{R.}~\bibnamefont{Loll}},
  \bibinfo{journal}{Nucl.Phys.} \textbf{\bibinfo{volume}{B536}},
  \bibinfo{pages}{407} (\bibinfo{year}{1998}), \eprint{hep-th/9805108}.

\bibitem[{\citenamefont{Ambjorn
  et~al.}(2005{\natexlab{c}})\citenamefont{Ambjorn, Jurkiewicz, and
  Loll}}]{Ambjorn:2005db}
\bibinfo{author}{\bibfnamefont{J.}~\bibnamefont{Ambjorn}},
  \bibinfo{author}{\bibfnamefont{J.}~\bibnamefont{Jurkiewicz}},
  \bibnamefont{and} \bibinfo{author}{\bibfnamefont{R.}~\bibnamefont{Loll}},
  \bibinfo{journal}{Phys. Rev. Lett.} \textbf{\bibinfo{volume}{95}},
  \bibinfo{pages}{171301} (\bibinfo{year}{2005}{\natexlab{c}}),
  \eprint{hep-th/0505113}.

\bibitem[{\citenamefont{Lauscher and Reuter}(2005)}]{Lauscher:2005qz}
\bibinfo{author}{\bibfnamefont{O.}~\bibnamefont{Lauscher}} \bibnamefont{and}
  \bibinfo{author}{\bibfnamefont{M.}~\bibnamefont{Reuter}},
  \bibinfo{journal}{JHEP} \textbf{\bibinfo{volume}{0510}}, \bibinfo{pages}{050}
  (\bibinfo{year}{2005}), \eprint{hep-th/0508202}.

\bibitem[{\citenamefont{Bilke et~al.}(1998)\citenamefont{Bilke, Burda,
  Krzywicki, Petersson, Tabaczek et~al.}}]{Bilke:1998vj}
\bibinfo{author}{\bibfnamefont{S.}~\bibnamefont{Bilke}},
  \bibinfo{author}{\bibfnamefont{Z.}~\bibnamefont{Burda}},
  \bibinfo{author}{\bibfnamefont{A.}~\bibnamefont{Krzywicki}},
  \bibinfo{author}{\bibfnamefont{B.}~\bibnamefont{Petersson}},
  \bibinfo{author}{\bibfnamefont{J.}~\bibnamefont{Tabaczek}},
  \bibnamefont{et~al.}, \bibinfo{journal}{Phys.Lett.}
  \textbf{\bibinfo{volume}{B432}}, \bibinfo{pages}{279} (\bibinfo{year}{1998}),
  \eprint{hep-lat/9804011}.

\bibitem[{\citenamefont{Bruegmann and Marinari}(1993)}]{Bruegmann:1992jk}
\bibinfo{author}{\bibfnamefont{B.}~\bibnamefont{Bruegmann}} \bibnamefont{and}
  \bibinfo{author}{\bibfnamefont{E.}~\bibnamefont{Marinari}},
  \bibinfo{journal}{Phys.Rev.Lett.} \textbf{\bibinfo{volume}{70}},
  \bibinfo{pages}{1908} (\bibinfo{year}{1993}), \eprint{hep-lat/9210002}.

\bibitem[{\citenamefont{Bilke and Thorleifsson}(1999)}]{Bilke:1998bn}
\bibinfo{author}{\bibfnamefont{S.}~\bibnamefont{Bilke}} \bibnamefont{and}
  \bibinfo{author}{\bibfnamefont{G.}~\bibnamefont{Thorleifsson}},
  \bibinfo{journal}{Phys.Rev.} \textbf{\bibinfo{volume}{D59}},
  \bibinfo{pages}{124008} (\bibinfo{year}{1999}), \eprint{hep-lat/9810049}.

\bibitem[{\citenamefont{Akkermans et~al.}(2010)\citenamefont{Akkermans, Dunne,
  and Teplyaev}}]{Akkermans:2010dz}
\bibinfo{author}{\bibfnamefont{E.}~\bibnamefont{Akkermans}},
  \bibinfo{author}{\bibfnamefont{G.~V.} \bibnamefont{Dunne}}, \bibnamefont{and}
  \bibinfo{author}{\bibfnamefont{A.}~\bibnamefont{Teplyaev}},
  \bibinfo{journal}{Phys.Rev.Lett.} \textbf{\bibinfo{volume}{105}},
  \bibinfo{pages}{230407} (\bibinfo{year}{2010}), \eprint{arXiv:1010.1148}.

\end{thebibliography}

\end{document}